\documentclass[11pt]{article}

\usepackage[most]{tcolorbox} 
\usepackage{capt-of}

\usepackage{adjustbox}
\usepackage{rotating}
\usepackage{multirow}
\usepackage{times}
\usepackage{graphicx}
\usepackage{color}
\usepackage{amsmath,bm}
\usepackage{lineno}
\usepackage{array}
\usepackage{delarray}
\usepackage{hyperref}
\usepackage[sort&compress]{natbib}

\begin{document}

\title{Statistical physics for artificial neural networks}




\author{Zongrui Pei$^{1}$ \\
\normalsize{$^{1}$New York University, New York, NY10003, USA}\\
\normalsize{E-mail: peizongrui@gmail.com; zp2137@nyu.edu} \\
}


\date{}

\baselineskip24pt

\maketitle 
 


{\it
{\bf Abstract}

{\color{black}
The 2024 Nobel Prize in Physics was awarded for pioneering contributions at the intersection of artificial neural networks (ANNs) and spin-glass physics, underscoring the profound connections between these fields. The topological similarities between ANNs and Ising-type models, such as the Sherrington-Kirkpatrick model, reveal shared structures that bridge statistical physics and machine learning. In this perspective, we explore how concepts and methods from statistical physics, particularly those related to glassy and disordered systems like spin glasses, are applied to the study and development of ANNs. We discuss the key differences, common features, and deep interconnections between spin glasses and neural networks while highlighting future directions for this interdisciplinary research. Special attention is given to the synergy between spin-glass studies and neural network advancements and the challenges that remain in statistical physics for ANNs. Finally, we examine the transformative role that quantum computing could play in addressing these challenges and propelling this research frontier forward.}

}

\tableofcontents

\section{Introduction}



\begin{figure}
    \centering
    \includegraphics[width=0.9\linewidth]{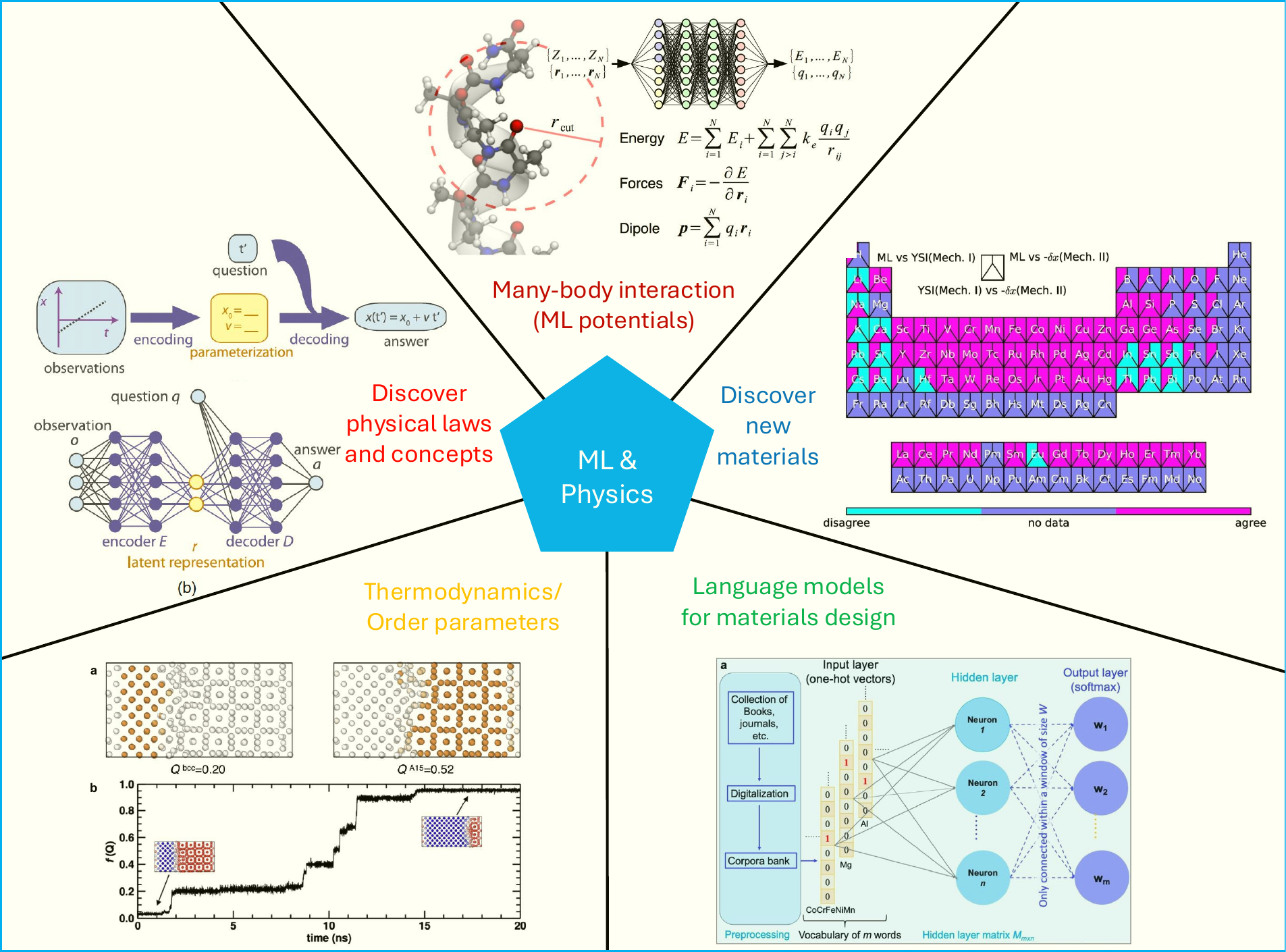}
    \caption{The applications of artificial neural networks in physics. Artificial neural networks have been widely utilized to solve a diverse range of problems, including those in scientific research. Here, we show a few typical examples in physical research, i.e., (i) many-body interactions (machine-learning potentials) \cite{unke2019physnet,zhang2018deep}, (ii) thermodynamics (proposal of order parameters to describe phase transitions) \cite{yin2021neural,rogal2019neural}, (iii) discovery of physical laws and concepts \cite{iten2020discovering}, (iv) discovery or design of new materials \cite{pei2021machine}, and (v) language models for materials design \cite{pei2023toward,Tshitoyan2019,pei2024towards}.
}
    \label{fig:applications}
\end{figure}

Artificial neural networks (ANNs) have had a profound influence in many sectors, as demonstrated by numerous notable milestones in ANN applications. For example, AlexNet is one of the first models to show the impressive capabilities of ANNs in classification tasks using the ImageNet dataset \cite{krizhevsky2012imagenet}. AlphaGo corroborates that ANNs can outperform human players in games \cite{silver2016mastering,silver2017mastering}. ANNs also find their application in self-driving cars \cite{badue2021self}. Recently, large language models \cite{chatGPT,pei2025language}, which are complex ANN models, have fundamentally changed how we work, learn, and teach, influencing nearly everyone.
ANNs have been widely adopted in various scientific research domains. An outstanding example is AlphaFold, which has solved the half-century-long challenge in structural biology \cite{jumper2021highly} and accelerated the determination of three-dimensional (3D) protein structures, potentially revolutionizing drug discovery and the healthcare industry. A few more typical examples in physics are shown in Figure \ref{fig:applications}. ANNs have been used to describe many-body interactions or construct empirical potentials, such as PhysNet \cite{unke2019physnet}. Some ANN potentials with multiple neural layers are also referred to as deep potentials \cite{zhang2018deep}. Recently, inspired by the success of foundation models for natural languages, developing foundation models for potentials has been proposed, and some initial studies have been conducted \cite{batatia2023foundation}. Order parameters can describe the phase transitions in thermodynamics. ANNs have been adopted to construct new order parameters successfully \cite{yin2021neural,rogal2019neural,jung2025roadmap}. As a proof of concept, ANNs are also used to rediscover physical laws and concepts \cite{iten2020discovering}.
ANN models have been developed to design novel materials \cite{pei2023toward,Tshitoyan2019,pei2024designing,pei2024computer}. They have been utilized to link the physical properties of various materials and their crystal structures, microstructure, and processing \cite{pei2021machine,pei2024towards}. Examples of applications include low-dimensional materials, structural materials, and functional materials, among others.

ANNs have been employed to understand Ising-like physical systems \cite{mills2020finding,carrasquilla2017machine,hibat2021variational,fan2023searching}, including spin glass \cite{hibat2021variational,fan2023searching}. They provide a new opportunity to understand the physics of spin glass. Fan {\it et al.} searched for spin glass ground states through deep reinforcement learning \cite{fan2023searching}. Huang {\it et al.} confirmed the efficiency of classic machine learning to describe quantum phases, taking 2D random Heisenberg models as an example \cite{huang2022provably}.
ANNs can boost Monte-Carlo simulations of spin glasses (autoregressive neural networks) \cite{mcnaughton2020boosting} and search for ground states of spin glasses (tropical tensor network) \cite{liu2021tropical}.

\begin{figure}
    \centering
    \includegraphics[width=1.0\linewidth]{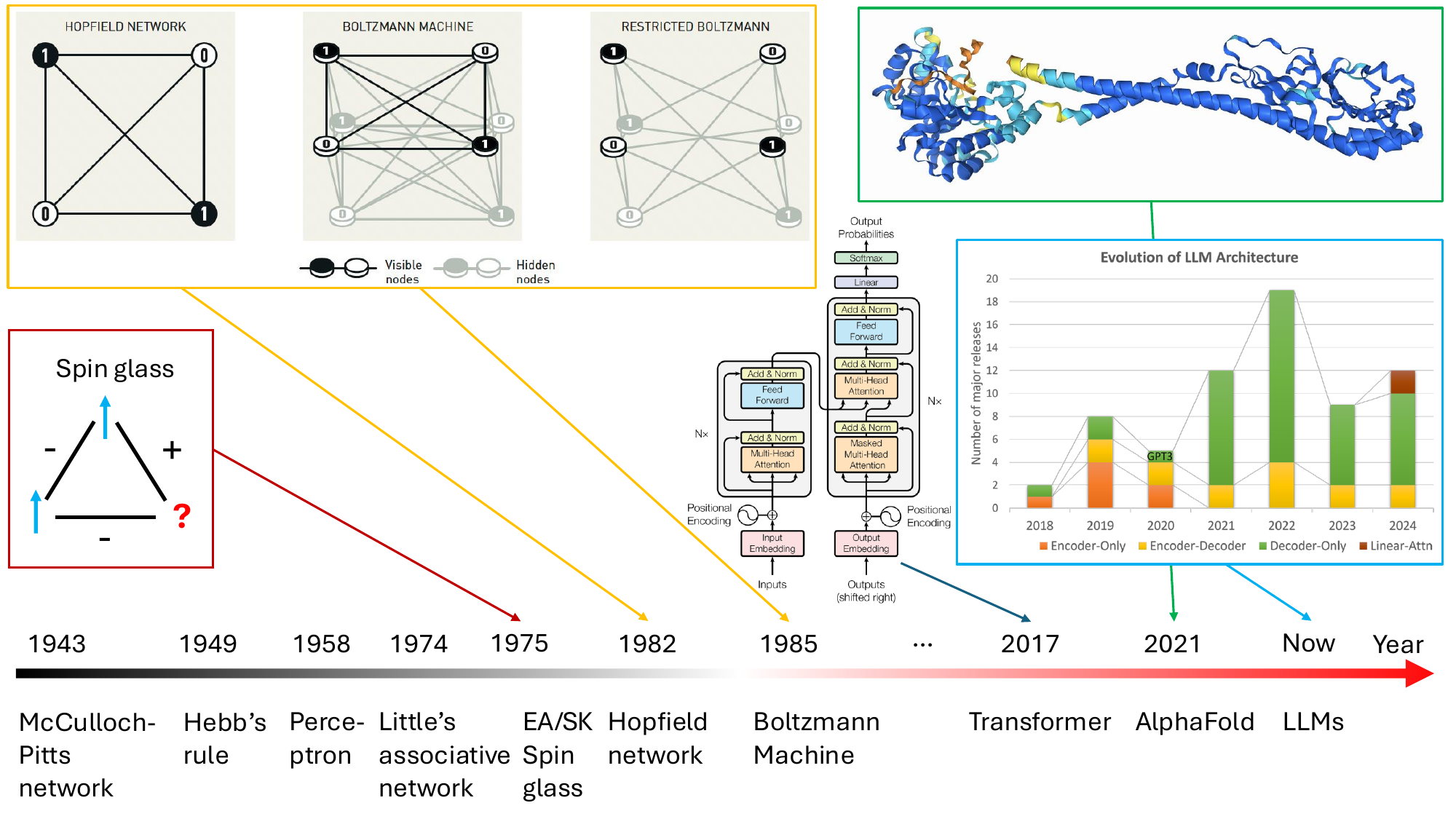}
    \caption{The history of artificial neural networks and relevant physical models. Other landmarks include backpropagation in the 1970s and AlexNet in 2012. Interestingly, bio-inspired ANNs ultimately led to the solution of the protein folding problem that has plagued us for half a century.}
    \label{fig:history}
\end{figure}

The significant breakthroughs in developing ANNs highlight the importance of convergence across different disciplines. ANNs mimic the structure and function of biological neurons, inspired by the human brain. In addition to their biological origin, the physical foundation of ANNs has been recognized by the Nobel Prize in Physics 2024, awarded to physicist John J. Hopfield and computer scientist Geoffrey E. Hinton \cite{hinton2025nobel}. Their foundational discoveries enabled the development of ANNs and their applications today. ANNs are closely connected with spin glass, a traditional topic theorized by Edwards and Anderson in 1975 \cite{edwards1975theory} and has attracted considerable attention in recent years. The importance of spin glasses as a representative complex system in different research domains across various length scales was demonstrated by the Nobel Prize in Physics 2021, awarded partly to Giorgio Parisi \cite{parisi1979infinite}. The Nobel Prizes in Physics in 2021 and 2024 have also contributed to the increased attention. The prize-winning studies revealed a profound connection between spin glasses and diverse disciplines, ranging from large to small scales, encompassing complexity science, biology, and computer science.

There are a few landmarks in the history of ANNs [Figure \ref{fig:history}], such as the seminal work of McCulloch and Pitts on neural networks in 1943 \cite{mcculloch1943logical} that initiated the research domain and the Hebbian mechanism or Hebb's rule in 1949 that suggests when neurons fire signals \cite{hebb2005organization}, the simple ANN perceptron proposed by Rosenblatt in 1958 \cite{rosenblatt1958perceptron}, and the proposal of associative neural networks by Little in 1974 \cite{little1974existence}, and later by Hopfield in 1982 \cite{hopfield1982neural}. In 1985, Hinton and colleagues proposed the Boltzmann machine \cite{ackley1985learning}, a stochastic recurrent neural network (RNN) where the energy state of each neural-network configuration follows the Boltzmann distribution. Restricted Boltzmann machines were proposed later by removing the connections between neurons within the same layer to improve the network's efficiency. The proposal of backpropagation in the 1970s, which Hinton and colleagues utilized to optimize ANNs like AlexNet in 2012, is another significant milestone. In 2017, the transformer architecture was proposed as an alternative to RNN structures, where the attention mechanism is a key novel component. Due to its high efficiency and suitability in large model optimization, it opened a new era for deep neural networks. AlphaFold \cite{jumper2021highly}, the Nobel-Prize-winning model, and the popular large language models like GPT-3.5/4/4.5 and Llama 2/3/4 all adopt the transformer architecture.

Albeit with enormous success, the mechanisms behind these successes remain enigmatic.
Fortunately, it has been demonstrated that idealized versions of these powerful networks are mathematically equivalent to older, simpler machine learning models, such as kernel machines \cite{belkin2018understand,jacot2018neural} or physical models \cite{jcrn-3nrc}. Therefore, the physical foundation of ANNs is closely related to magnetism (spins) and statistical physics \cite{amit1985spin,amit1987statistical,watkin1993statistical,mezard2024spin}. 
Hopfield networks \cite{hopfield1982neural,hopfield1999brain,krotov2023new} and Boltzmann machines \cite{ackley1985learning} are essentially Sherrington-Kirkpatrick (SK) spin glasses with random interaction parameters \cite{sherrington1975solvable}, where each neuron or spin is connected with all other ones except itself.
When mapping the spin-glass structure to an ANN, each state corresponds to a pattern or memory in the Hopfield network. Finite metastable states indicate the limited capability of the Hopfield network \cite{folli2017maximum}.
It is essential to determine whether general spin glasses exhibit infinite metastable states, as is the case with the mean-field SK model.
Spherical spin glass models have been used to study simplified, idealized ANNs \cite{choromanska2015loss}.  Parisi proposed a replica symmetry breaking (RSB) solution \cite{parisi1979infinite,charbonneau2023spin} to the SK model \cite{sherrington1975solvable}. The RSB theory has also been used to study ANNs, and steps toward some rigorous results were discussed \cite{agliari2020replica}. Ghio {\it et al.} showed the sampling with flows, diffusion, and autoregressive neural networks from a spin-glass perspective \cite{ghio2024sampling}.  If this equivalence of ANNs and spin-glass systems can be extended beyond idealized neural networks, it may explain how practical ANNs achieve their astonishing results.

\newpage

\tcbset{enhanced,colback=cyan!5!white,colframe=cyan!75!black,fonttitle=\bfseries,fontupper=\linespread{1.5}\selectfont}

\begin{tcolorbox}[breakable, every float=\centering, drop shadow, title=Box 1 \textbar\ Basic Concepts of Statistical Physics Used by ANNs,width=\dimexpr\textwidth+12mm\relax,enlarge left by=-6mm, boxsep=5pt]


\section*{Thermodynamics}


{\bf Energy Landscapes and Loss Functions}
The loss functions in ANNs are similar to the total energy of physical systems [Figure \ref{fig:basic-concepts}]. Statistical physics concepts, such as energy landscapes, are used to understand the dynamics and optimization in neural networks. For example, the weights and states in neural networks can be thought of as occupying a rugged energy landscape with many local minima, just like glassy Ising models.
To optimize or train ANNs, we use methods such as simulated annealing and stochastic gradient descent, which are inspired by statistical physics.


{\bf Entropy and Information Theory}
Entropy from statistical physics is used to measure uncertainty in predictions and model behavior. One type of entropy, known as "cross entropy," between the actual values and predictions can be used as an error function. In addition, the entropy concept is closely related to the Information Bottleneck Principle, which explains how neural networks compress information during training, much like physical systems reduce entropy under certain conditions.

\begin{minipage}[t]{1.0\linewidth}
    \vspace*{0pt}
    \centering
    \includegraphics[width=0.66\linewidth]{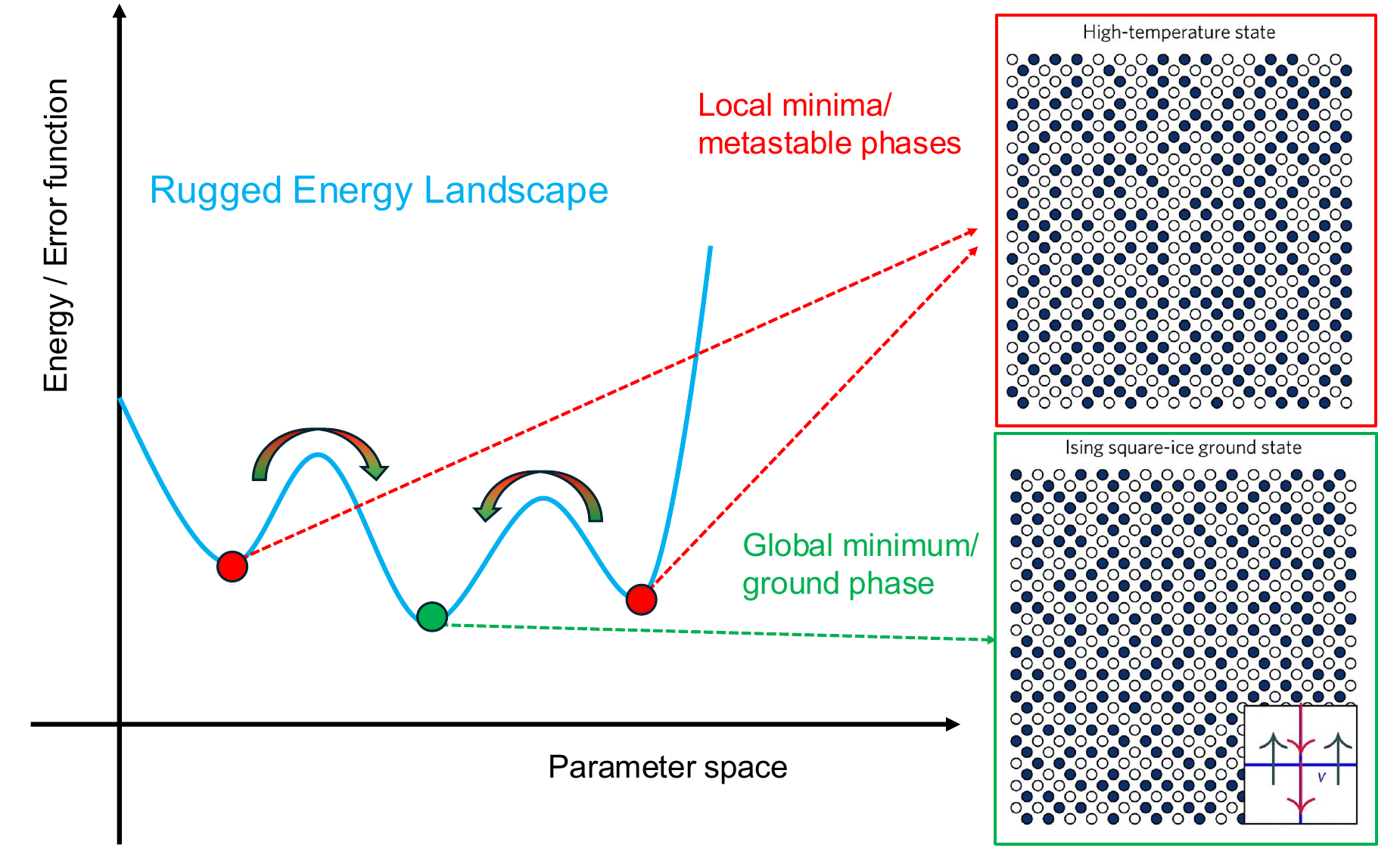}
    \captionof{figure}{Basic concepts in statistical physics and thermodynamics. Here, we show that each pattern (memory or configuration) corresponds to an energy state in the rugged energy landscape of the free energy function or an error function.\\}
    \label{fig:basic-concepts}
\end{minipage}

{\bf Phase Transitions}
The concept of phase transitions is useful to describe the behavior of ANNs. ANNs exhibit phase transitions during training when their prediction and reliability suddenly change upon small changes of ANN structures (e.g., neural layers, parameter size in each layer), learning rate, etc., around critical values. This is similar to physical systems undergoing structural changes (e.g., spin glasses). 


\section*{Bayesian Inference and Partition Functions}
Bayesian statistics finds its application in ANNs. It uses Bayes' theorem to calculate and update the probability distribution when new data or knowledge is available. This conditional feature meets the flexibility requirement of real-world problems, making it practically useful. Bayesian neural networks use principles from statistical physics for probabilistic inference. The partition function, a cornerstone of statistical mechanics, is used to calculate probabilities in probabilistic graphical models, including some neural networks.

\section*{Learning Dynamics and Langevin Equations}
There are a few connections between Langevin equations and ANNs. Here, we mention two of them: 
(i) The learning dynamics of ANNs is described by Langevin equations, analogous to the motion of particles in a thermal bath. In studying stochastic gradient descent (SGD), the noise introduced by mini-batch sampling is found to resemble thermal noise \cite{goldt2019dynamics}.
(ii) The diffusion model of ANNs for images is similar to Langevin equations \cite{albergo2023stochastic}, since both start with a white noisy background, and gradually recover the true states based on the status change of each particle or pixel.

\end{tcolorbox}



\section{Spin-glass physics for artificial neural networks}


Just as neurons are the biological analogue of ANNs, spin glasses could be considered the physical analogue of ANNs: they were the inspiration for the original Hopfield model~\cite{hopfield1982neural}, which itself is viewed by some as a type of spin glass. 
We will discuss their relationships below.

\subsection{Relations of spin glasses, biological neurons, and associative memory}

{\bf Spin glass}
A spin glass is a type of magnetic system in which the quenched interactions between localized spins may be either ferromagnetic or antiferromagnetic with roughly equal probability. It exhibits an array of experimental properties: spin freezing with no long-range magnetic order at low temperature, a magnetic susceptibility cusp at a critical temperature accompanied by the absence of a singularity in the specific heat, slow relaxation, aging, and several others~\cite{stein2013spin,stein2011spin}. 
Most theoretical work has focused on the Edwards-Anderson (EA) spin glass~\cite{edwards1975theory} proposed in 1975 and its infinite-range analogue, the Sherrington-Kirkpatrick (SK) model~\cite{sherrington1975solvable}, proposed the same year. 

The EA Hamiltonian in the absence of an external field is given by
\begin{equation}
\label{eq:H-SG}
H=-\sum_{<ij>} J_{ij} \sigma_i \sigma_j\ ,
\end{equation}
where the couplings $J_{ij}$ are i.i.d.~random variables representing the interaction between spins $\sigma_i$ and $\sigma_j$ at nearest-neighbor sites $i$ and $j$ (denoted by the bracket in~(\ref{eq:H-SG})). One is free to choose the distribution of the couplings: a common choice, which will be used here, is a Gaussian distribution with zero mean and unit variance. This distinguishes the spin glass from more conventional magnets: if $J_{ij}$ were a constant $J$, the model becomes either a ferromagnet ($J>0$) or an antiferromagnet ($J<0$), both with an ordered ground state. 

As noted above, when every pair of spins interacts with each other, the model becomes the SK~model, a mean-field spin glass with Hamiltonian
\begin{equation}
\label{eq:SK}
H=-\frac{1}{\sqrt{N}}\sum_{i<j} J_{ij} \sigma_i \sigma_j\ ,
\end{equation}
where the coupling distribution $J_{ij}$ is also a mean-zero, unit variance Gaussian. The thermodynamic properties of the SK model are well understood, and analytical expressions were found for its free energy and order parameters \cite{parisi1979infinite}. Whether similar properties exist in short-range spin glasses remains an open question. 
Understanding short-range spin glasses in finite dimensions remains a major challenge in both condensed matter physics and statistical mechanics.

{\bf ANNs and spin glasses} 
Magnetic systems have long been used to model neural networks, going back (at least) to the work of McCullough and Pitts~\cite{mcculloch1943logical}. In its simplest form, a neuron is assumed to have only two states: `on' or firing, and `off' or quiescent. Therefore, its state can be modelled as an Ising spin, where $+1$ corresponds to the firing state and $-1$ to the quiescent state. A state, or firing pattern, of the entire neural network then corresponds to a spin configuration $\{\sigma\}$. The interactions between neurons, i.e., synaptic efficiencies, correspond to the couplings $J_{ij}$ between spins in a magnetic system.

The Hebb learning rule is used to determine the coupling parameters \cite{hebb2005organization}.
In associative neural networks, such as the Hopfield network, suppose there are $p$ patterns (corresponding to memories).
Then the interactions $J_{ij}$ between neurons are modelled as
\begin{equation}
J_{ij} =\frac{1}{N} \sum_{\mu=1}^p \xi_i^{\mu} \xi_j^{\mu},
\end{equation}
where the state of the $i^{\rm th}$ spin in the $\mu^{\rm th}$ pattern (spin configuration) is represented by $\xi_i^{\mu}$, which also takes on the values $\pm 1$.  
When $p=1$, the system is equivalent to a Mattis model~\cite{mattis1976solvable}, which is gauge-equivalent to a ferromagnet. Setting $p>1$ introduces frustration into the system, and its behavior becomes more spin-glass-like. 
Apart from its interpretation as a model for neural networks, the Hopfield model represents an interesting statistical mechanical system in its own right and has been the subject of considerable study. A few studies have used statistical mechanics methods to investigate phase transitions in the Hopfield model \cite{hopfield1982neural,amit1985storing,gardner1988space} and have found a bound $p$ on the number of memories that can be faithfully recalled from any initial state in a system of $N$ spins or neurons.

\begin{figure}
    \centering
    \includegraphics[width=1.0\linewidth]{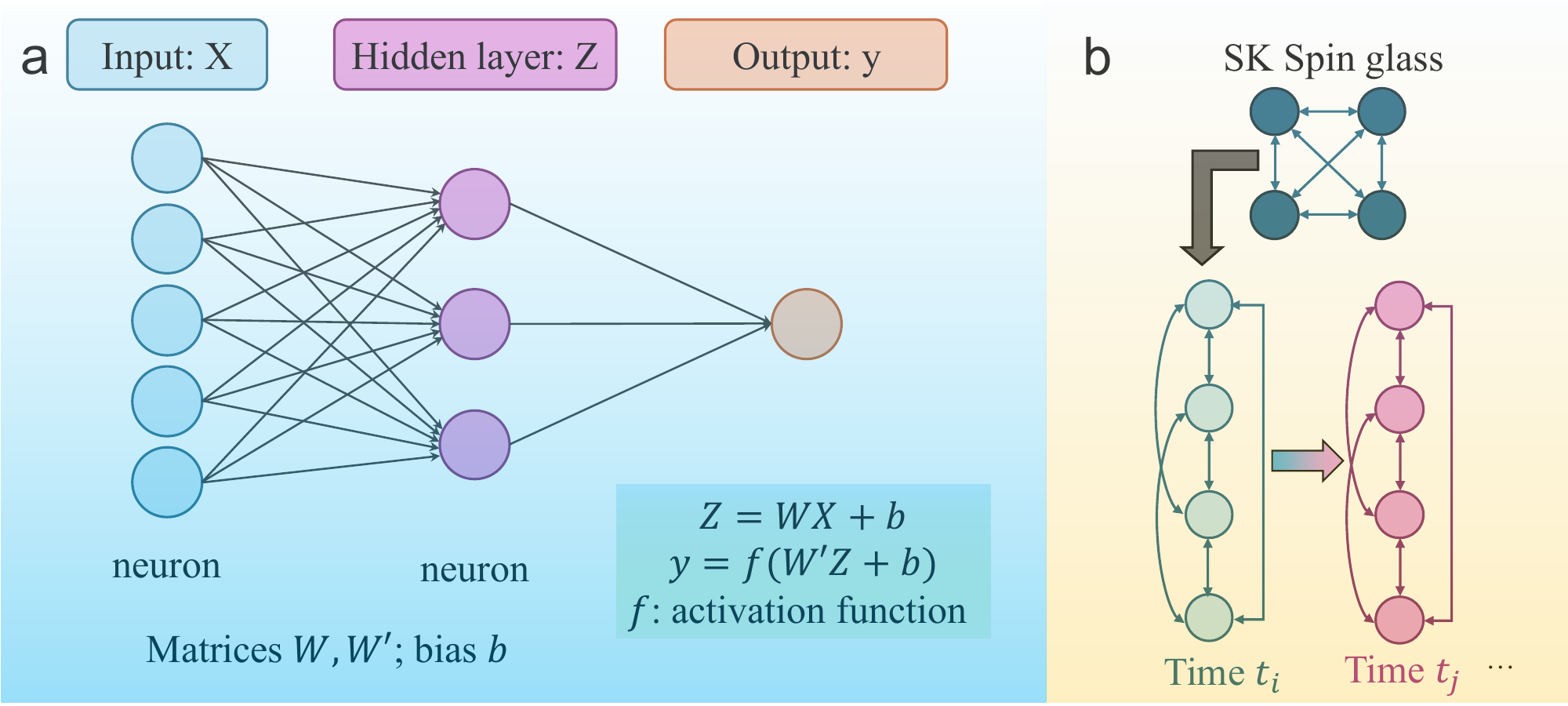}
    \caption{Topology of an artificial neural network and Sherrington-Kirkpatrick (SK) model. {\bf a}, A simple, fully connected neural network with one hidden layer. {\bf b}, A four-spin mean field SK spin-glass model.}
    \label{fig:Figure2}
\end{figure}

In addition to Hopfield networks, some idealized ANNs also have similar correspondences with spin glasses~\cite{little1974existence}, such as the spherical SK spin glass models~\cite{choromanska2015loss} [Figure \ref{fig:Figure2}]. 
In a spherical spin glass, the spins have varying magnitudes subject to the constraint $\sum_i^N\sigma_i^2=N$.
Given training data $\{X_i\}, y$, a neural network with $H-1$ hidden layers is equivalent to an equation \cite{choromanska2015loss}, i.e., $y = q \sigma\big{(}W^T_H \sigma(W^T_{H-1} ... \sigma(W_1^T X))...\big{)}$,
where $q$ is a normalization factor.
The loss function of this ANN is 
\begin{equation}
L_{\Lambda, H}(\bm{\tilde{w}}) = \frac{1}{\Lambda^{(H-1)/2}} \sum_{i_1,i_2,...,i_H=1}^{\Lambda} X_{i_1,i_2,...,i_H} \tilde{w}_{i_1} \tilde{w}_{i_2} ...\tilde{w}_{i_H},
\end{equation}
where $\Lambda$ is the number of weights.
By imposing the spherical constraint above on the weights that follow $1/\Lambda \sum_{i=1}^{\Lambda} \tilde{w}_i^2 =1$, the loss function of this neural network was found to be mathematically equivalent to the Hamiltonian formulation of a spin glass.


\begin{figure}
    \centering
    \includegraphics[width=1.0\linewidth]{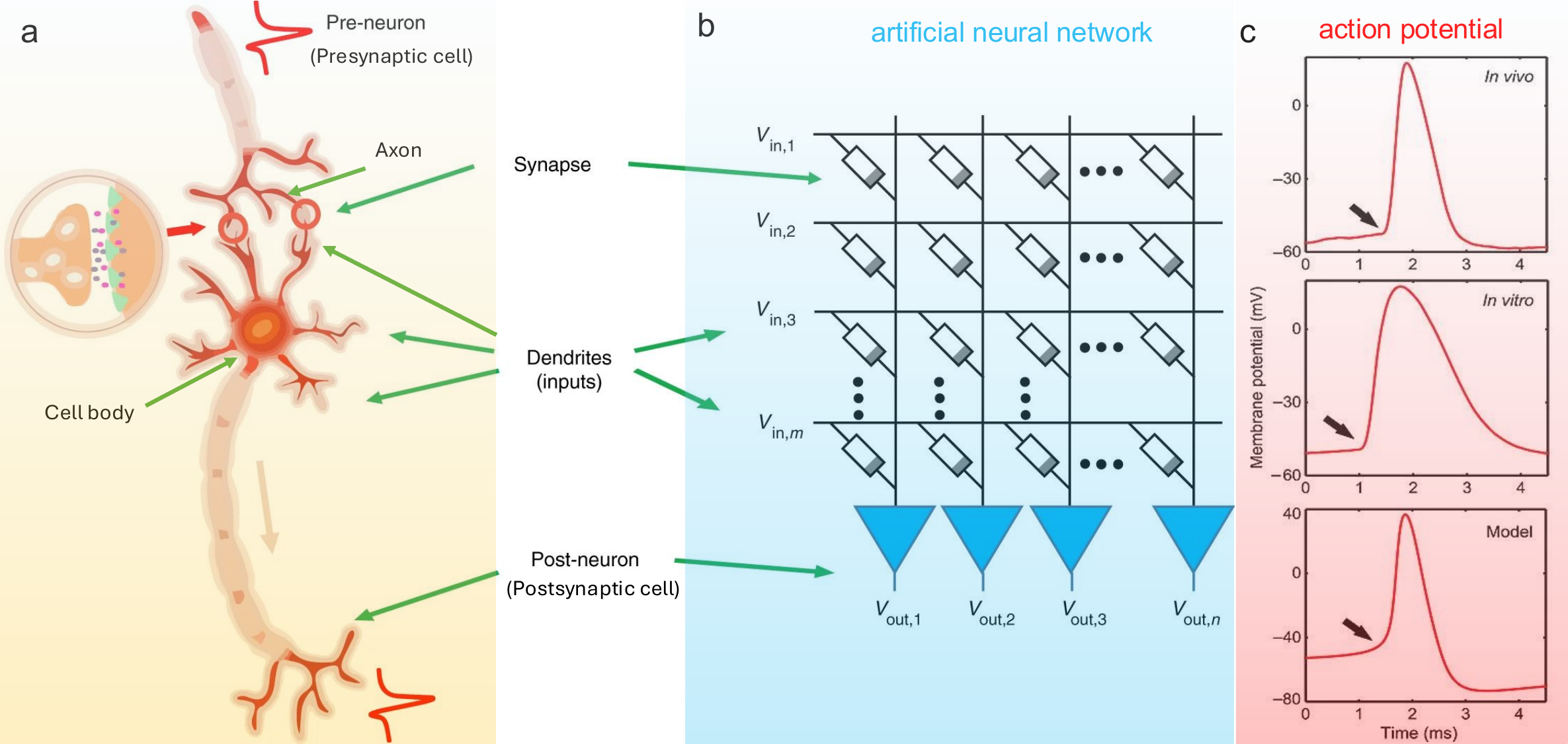}
    \caption{Biological neurons, artificial neural networks, and action potentials. {\bf a}, Biological neuron and its associated concepts \cite{duan2020spiking}. {\bf b}, Artificial neural network with input features $V_{\mathrm{in},i}$ and output features $V_{\mathrm{out},i}$ \cite{duan2020spiking}. Here, the resistor symbols (rectangles) represent the activation functions that switch the signals from individual inputs on or off. {\bf c}  Activation potentials \cite{naundorf2006unique}. The top panel represents an action potential in a cat visual cortex neuron {\it in vivo}. The middle panel is an action potential from a cat visual cortical slice {\it in vivo} at 20°C. The bottom panel is a model potential. The arrow indicates the characteristic kink at the onset of the action potential. This figure is adapted from Refs. \cite{duan2020spiking,naundorf2006unique} and modified. }
    \label{fig:bio-neuro-ANNs}
\end{figure}

{\bf Similarity between biological neurons and ANNs~}
ANNs are designed to imitate the structure of biological neurons [Figure \ref{fig:bio-neuro-ANNs}] \cite{duan2020spiking}. A neuron or nerve cell consists of three major parts, i.e., a cell body (soma), dendrites (receiving extensions), and an axon (conducting extension). When two neurons are connected, the presynaptic cell sends an electromagnetic signal to the postsynaptic cell. The signal travels from the axon of the presynaptic cell to the dendrites of the postsynaptic cell, which are directly connected. A threshold potential or action potential controls the activation of a neural connection [Figure \ref{fig:bio-neuro-ANNs} {\bf c}]. The postsynaptic potential decides when to fire an electromagnetic signal. The potential at a specific neuron is a function of all postsynaptic potentials from its presynaptic neurons, i.e., 
\begin{equation}
V_i =\sum_j J_{ij}(S_j +1).
\end{equation}
When $V_i$ is larger than a threshold $U_i$, or $V_i-U_i >0$ the neuron is active ($S_i=1$). Using the Heaviside function $H(x)$, the state of the neuron can be written as $S_i=H(U_i - V_i)$. We use $h_i$ to represent the molecular field, $h_i=U_i - V_i$. When the spin direction is parallel to $h_i$, the local configuration is stable, i.e., $h_i S_i >0$.
Usually, the threshold function $U_i$ is assumed to satisfy $U_i \approx \sum_j J_{ij}$, then we find $-\frac{1}{2}\sum_i h_i S_i \approx H =-\frac{1}{2}\sum_{ij} J_{ij} \sigma_i \sigma_j$, which has the same form as Eq. \ref{eq:H-SG}.
The above formulation is for zero temperature. At finite temperature, the probability of activating one neuron is $P(S_i) = \frac{1}{\exp [-\beta (V_i-U_i)] +1}$,
which is a Fermi-Dirac distribution. Here, $\beta$ is the inverse temperature, defined by $\beta=1/k_B T$ with $k_B$ as the Boltzmann constant and $T$ is the temperature.

ANNs inherit these key features. Each artificial neuron sends data to the neurons in its following layers, which react collectively. This response is calculated using matrix multiplication plus a bias variable. The activation and updated connections of neurons follow Hebb's rule~\cite{hebb2005organization}, which states that when two neurons fire together, the excitatory (ferromagnetic in magnetic terms) component of their coupling is enhanced. The more often two neurons are active together, the stronger their excitatory connection becomes. This holds for both biological and artificial neural networks.

\subsection{Dictionary of corresponding concepts}

To clarify the connection across the three research domains, we provide a list of analogous features of ANNs, biological neurons, and spin glasses in Table~\ref{tab:dictionary-ANN-SG}. Here, the concepts across the three research domains are compared. A few concepts in the table have been discussed above, such as a spin in spin glass corresponds to a biological neuron in biology and an artificial neuron in ANNs. This comparison helps identify some interesting concepts that are not studied, thus deepening our understanding and providing new research opportunities. For example, the Hamiltonian in spin glasses yields a total energy of the system that has a similar role to the loss function in ANNs, which has no correspondence in biological neurons.  
The effective learning mechanism of neurons remains elusive, although a few explanations have been proposed, such as the principle of predictive coding \cite{luczak2022neurons}.
The concept of Hamiltonian, if it can capture the overall activity of biological neurons, may be relevant to what dominates the coordination of specific neurons and the choice of particular information propagation paths \cite{friston2010free,friston2009free}. Various forms of order parameters, including spin overlap parameters, magnetization, giant cluster size (commonly used in percolation theory), etc., are physical quantities that describe the average information of spin variables in spin systems and capture phase states. It can be calculated as the overall activity of biological neurons or artificial neurons, like how closely the current firing pattern matches a learned activity pattern in the hippocampus or cortex. Studying order parameters in ANNs can help understand how many neurons are generally used or activated, which is valuable to understanding the fundamental principles of ANNs as kernels. 

\begin{table}[]
    \centering    
    \caption{Dictionary for artificial neural networks (ANNs), spin glasses, and biological neurons. The terms in spin glasses and their corresponding terms in ANNs are compared. Interestingly, no term in ANNs corresponds to the order parameter, which is closely connected to the value of loss functions. The common features of ANNs and spin glasses explain (i) why Monte Carlo methods can explore their energy landscapes and (ii) why spin glass methods can be directly used to study ANNs. Another interesting question is whether we need to introduce new order parameters for spin glass physics. }
    \begin{tabular}{lll}
    \hline \hline
    spin glass & biological neuron & artificial neural network \\
       \hline
        spin & biological neuron & artificial neuron  \\
        spin variables & neuron state (active/inactive) & features \\
        interaction of two spins & electromagnetic signal & weight between two neurons \\
        interaction strengths & action potential & weight matrix \\
        total energy (Hamiltonian) &? (unclear) & loss function \\
        stable or metastable states & memory & memorized patterns in associative network \\
        connectivity of spins & synapses & activation function\\
        spherical SG models &? (unknown) & topology of fully connected ANNs\\
        order parameter & overall activity of neurons & ? (no correspondence) \\
    \hline \hline
    \end{tabular}

    \label{tab:dictionary-ANN-SG}
\end{table}

\subsection{Hopfield neural network and Boltzmann machines}

The Hopfield neural network \cite{hopfield1982neural} consists of a single layer of fully interconnected neurons, with each neuron linked to every other neuron [Figure \ref{fig:Figure2}{\bf a}]. The metastable states, or local minima of the network, correspond to the patterns, which allows the creation of associative or content addressable memories. Patterns are memorized and encoded in the network parameters. A similar idea was proposed by Little eight years prior to Hopfield's discovery \cite{little1974existence}. Structurally, it bears a topological resemblance to the SK model \cite{sherrington1975solvable}. The phase diagram of the Hopfield network is also similar to the SK model \cite{sherrington1975solvable}. The original motivation of the Hopfield network was to create a model of associative memory, in which a stored pattern can be retrieved from an incomplete or noisy input. 

The Hopfield network evolves over time, with the state of each neuron changing dynamically (Figure \ref{fig:Figure2}b). This temporal evolution allows it to be viewed as a multi-layered system, where each “layer” represents a different time step. In this sense, it functions as a fully connected neural network, with each neuron capable of reading and outputting data. The network stores patterns by adjusting connection weights, and the number of patterns it can retain is proportional to the number of neurons \cite{folli2017maximum}.

The Hopfield neural network influenced the development of recurrent neural networks (RNNs), such as the long short-term memory networks and the more efficient gated recurrent units. RNNs, characterized by recurrent neural layers, are designed to process sequential data, such as speech and natural languages. The transformer architecture has gradually replaced these methods, where the self-attention mechanism is adopted to replace recurrence for processing sequential data. The Hopfield neural network and its successors remain an active area of research~\cite{agliari2019relativistic,saccone2022direct,negri2023storage}.

Similar to the Hopfield neural network, the Boltzmann machine is a system where each spin interacts with all the others \cite{ackley1985learning} [Figure \ref{fig:figure3}{\bf b}]. It consists of a visible layer and a hidden layer, with all neurons—whether in the visible (input/output) or hidden layer—fully connected. Data is both inputted and outputted through the visible layer. A defining characteristic of Boltzmann machines is their stochastic nature: neuron activation is probabilistic rather than deterministic, influenced by connection weights and inputs. The probability of activation is determined by the Boltzmann distribution. 

Due to the computational complexity of Boltzmann machines, more tractable ``restricted" Boltzmann machines (RBMs) were introduced \cite{smolensky1986information,nair2010rectified} [Figure \ref{fig:figure3}{\bf c}]. 
RBMs adopt a bipartite structure, where the digital layer (or visible layer) and the analog layer (hidden layer) are fully connected, but there are no intra-layer connections.
This structural constraint allows the use of more computationally efficient training algorithms compared to RBMs \cite{fischer2014training}. Boltzmann machines have been widely applied to physical and chemical problems, including quantum many-body wavefunction simulations \cite{nomura2017restricted,melko2019restricted}, modeling polymer conformational properties \cite{yu2019generating}, and representing quantum states with non-Abelian or anyonic symmetries \cite{vieijra2020restricted}.


Barra {\it et al.} studied the equivalence of Hopfield networks and Boltzmann machines (Figure \ref{fig:figure3}{\bf d}) \cite{barra2012equivalence}. The study is based on a ``hybrid" Boltzmann machine (HBM) model, where the $P$ hidden units in the analog layer are continuous (for pattern storage) and the $N$ visible neurons are discrete and binary. They showed that the HBM, when marginalized over the hidden units, and the Hopfield network are statistically equivalent. 
Assume $P(\sigma,z)$ is the joint distribution for HBM, $P(z)$ is the distribution for the continuous hidden variables that usually follow the Gaussian distribution, and 
$P(\sigma)$ is the distribution for the Hopfield distribution. These quantities follow Bayes' rule $P(\sigma,z)=P(\sigma |z)P(z)=P(z|\sigma)P(\sigma)$.

Since this work involves several key concepts used in ANNs, such as the diffusion model and pattern overlap, we discuss their theoretical details further here.
The activity of the hidden layer follows a stochastic differential equation $T\frac{dz_{\mu}}{dt} =-z_{\mu} (t) + \sum_i \xi_i^{\mu} \sigma_i + \frac{2T}{\beta} \zeta_{\mu}(t)$,
where $\zeta_{\mu}$ is a white Gaussian noise. The idea is similar to the diffusion model widely used in image and video generation.
The probability for $z_{\mu}$ described by the stochastic differential equation above is $P(z_{\mu}|\sigma) = \sqrt{\frac{\beta}{2\pi}} \exp \bigg{[} -\frac{\beta}{2} \bigg{(}z_{\mu} - \sum_i \xi_i^{\mu} \sigma_i \bigg{)}^2 \bigg{]}$.
The Hamiltonian of the HBM shown in Figure \ref{fig:figure3}{\bf d} is $H_{hbm} (\sigma,z,\tau;\xi,\eta) = \frac{1}{2} \bigg{(}\sum_{\mu} z^2_{\mu} + \sum_{\nu} \tau^2_{\nu}\bigg{)} -\sum_i \sigma^2_i \bigg{(}\sum_{\mu} \xi_i^{\mu} z_{\mu} + \sum_{\nu} \eta_i^{\nu} \tau_{\nu} \bigg{)}$.
Then, the joint probability for the HBM can be calculated by
$P(\sigma,z,\tau) = \exp[-\beta H_{hbm} (\sigma,z,\tau;\xi,\eta)]/Z(\beta,\xi,\eta)$,
where the partition function
$Z(\beta,\xi,\eta) =\sum_{\sigma} \int \prod^P_{\mu=1} dz_{\mu} \int \prod^K_{\nu=1} d\tau_{\nu} \exp[-\beta H_{hbm} (\sigma,z,\tau;\xi,\eta)]$.
Assisted by the Gaussian integral and Bayes' rule, the probability for $\sigma$ is $P(\sigma) = \exp \bigg{(}-\frac{\beta}{2} \sum_{i,j} (\sum_{\mu} \xi_i^{\mu} \xi_j^{\mu}) \sigma_i \sigma_j \bigg{)}$.
If we set $J_{ij} = \sum_{\mu} \xi_i^{\mu} \xi_j^{\mu}$, it is straightforward to see that this is the SK spin glass model.
After determining the HBM Hamiltonian, it is not difficult to find that the concept of pattern overlap is mathematically equivalent to the overlap of replicas, an order parameter used in spin glass \cite{edwards1975theory}.

Both Boltzmann machines and Hopfield neural networks have been foundational in the advancement of ANNs and deep learning. While newer, more efficient algorithms continue to emerge, these models demonstrate the profound impact of statistical physics on shaping machine-learning techniques.

\subsection{Replica theory and the cavity method}

{\bf Replica theory}
Replicas are copies of the same system. 
The key idea is to consider $n$ replicas in calculating its free energy, which is given by
\begin{equation}
F= k_B T \langle \ln(Z) \rangle,
\end{equation}
where $\langle \cdot \rangle$ represents thermal average over the ensemble.
Then, the mathematical identity 
\begin{equation}
\langle \ln(Z) \rangle =\lim_{n\rightarrow 0} \frac{\langle Z^n \rangle-1}{n}
\end{equation}
can be used at the limit of $n\rightarrow 0$ to find the free energy. This is purely mathematical and not physical because the initially assumed integer $n$ is treated like a real number that can be smaller than one and infinitely close to 0. Mathematician Talagrand proved its correctness strictly \cite{talagrand2003spin}.
In the replica symmetry method, each replica is treated identically, which is the origin of the negative entropy in the mean-field solution for the SK model. Parisi proposed a replica symmetry-breaking method that successfully addressed the negative entropy issue by constructing a matrix ansatz in which replicas can have different ordering states.

The replica theory was initially developed to study glassy, disordered systems, such as spin glasses \cite{kirkpatrick1978infinite,parisi1983order,charbonneau2023spin,newman2024critical}, and later to understand the macroscopic behavior of learning algorithms and capacity limits \cite{gardner1988space,amit1985storing}. It was used to analyze the phase transitions of associative neural networks (e.g., the Hopfield network) and overparameterization and generalization of ANNs \cite{rocks2022memorizing,baldassi2022learning}. 
One crucial question in associative neural networks is the storage capacity of memory. A network system with size $N$ could only provide associative memory $p \le \alpha_c N$ at zero temperature with $\alpha_c=0.1-0.2$ for Hopfield models \cite{hopfield1982neural,amit1985storing}. For example, Amit and colleagues found $\alpha_c \approx 0.138$ with Hebb's rule $J_{ij}=1/2\sum_{\mu} \xi_i^{\mu} \xi_j^{\mu}$ \cite{amit1985storing}. When different or no constraints are imposed on $J_{ij}$, different values of $\alpha_c$ are found \cite{gardner1988space,gardner1987maximum,gardner1988optimal,krauth1989storage}. When $p \ge \alpha_c N$, memory quality degrades quickly. There is a phase transition at $\alpha_c$. Studying this problem is equivalent to finding the number of ground states of a spin glass.

The replica theory has been used to understand the phase transition or the critical behavior of ANNs in supervised and unsupervised learning. Hou {\it et al.} proposed a statistical physics model of unsupervised learning and found that the sensory inputs drive a series of continuous phase transitions related to spontaneous intrinsic-symmetry breaking \cite{hou2020statistical}.
Baldassi {\it et al.} studied the subdominant dense clusters in ANNs with discrete synapses \cite{baldassi2015subdominant}. They found these clusters enabled high computational performance in these systems.

{\color{black}
{\bf Cavity method} Later, Parisi and coauthors proposed another method, i.e., the cavity model, to solve the SK model (a method to calculate the statistical properties by removing a single spin and observing the reaction of the spin system) \cite{mezard1987spin}. It is a statistical method used to calculate thermodynamic properties, serving as an alternative to the replica theory. It focuses on removing one spin and its interactions with its neighbors to create a cavity and calculates the response of the rest of the system. At zero temperature, the response is determined by energy minimization. Rocks {\it et al.} adopted the ``zero-temperature cavity method" for the random nonlinear features model \cite{rocks2022memorizing} to study the double descent behavior, an important phenomenon that we will discuss later.
}

\begin{figure}
    \centering
    \includegraphics[width=1.0\linewidth]{Figure3-v2.pdf}
    \caption{Comparison of Hopfield neural network and Boltzmann machines. {\bf a}, Principle of the Hopfield neural network. {\bf b}, Illustration of a Boltzmann machine. {\bf c}, Restricted Boltzmann machine, generated by removing the intra-layer connections of visible and hidden layers. {\bf d}, The equivalence of Hopfield neural network and a hybrid restricted Boltzmann machine. Here, the visible variables $\{\sigma_i\}$ are discrete and binary and hidden variables $\{z_i\}$ and $\{\tau_i\}$ are continuous.} 
    \label{fig:figure3}
\end{figure}

\subsection{Overparameterization and double-descent behavior}
{\bf Overparameterization} The variance-bias trade-off is prevalent and has been observed in numerous models, and the reason is apparent: fewer parameters result in low variance and high bias, while more parameters lead to high variance and low bias for each prediction. There is an optimal choice of intermediate parameter size at which the model achieves its best performance. However, this does not seem to hold for neural networks. The optimal performance of neural networks is achieved with overparameterization \cite{belkin2019reconciling,rocks2022memorizing,baldassi2022learning}.
When a neural network has more parameters than the number of input data points, it is still considered overparameterized. In simple analytical expressions, such as linear equations, this typically becomes problematic, a phenomenon known as overfitting.  
One long-standing mystery of ANNs is that they seemingly subvert traditional machine learning theory, as their parameter size can exceed the number of data points for training without a signal of overfitting.

{\bf Double-descent behavior} 
The total error function of a model usually has a ``U" shape, and its minimum corresponds to the optimal parameters. The training error vanishes at a critical value of the parameter size when it equals the size of the training data points. The test error becomes divergent at this critical parameter size, analogous to the specific heat at the transition temperature. In traditional linear models, the error increases except at this critical value; in the overparameterized model, the test error first decreases, then increases at the critical value, and then decreases again. This phenomenon is referred to as double-descent behavior \cite{belkin2019reconciling,rocks2022memorizing,baldassi2022learning}. This behavior results in an optimal or minimal test error in the overparameterized region. 

Understanding the variance-bias trade-off of neural networks with overparameterization is crucial in deep learning. 
There are different opinions on the overparameterization phenomenon, which relies on methods from statistics and probability theory. For example, this behavior can be studied using the replica trick. The neural networks undergo a phase transition when the parameters increase across the critical threshold.
The Information Bottleneck Principle (IBP) proposes that deep neural networks first fit the training data and then discard irrelevant information by going through an information bottleneck, which helps them generalize \cite{tishby2000information,tishby2015deep}. Since the results are based on a particular type of ANNs, it is controversial whether the IBP conclusion holds generally for all deep neural networks \cite{saxe2019information}.

ANNs with more neurons in a neural layer or wide neural layers usually have better generalization performance than their narrower counterparts. When the number of neurons approaches infinity, this extreme case becomes mathematically more straightforward to treat, which is equivalent to the fact that the SK model is easier to solve than the EA model. When an infinite number of neurons is allowed in a layer, this is equivalent to a Gaussian process \cite{neal1996priors,bahri2024houches}. Interestingly, a recent study came to similar conclusions for quantum neural networks \cite{garcia2025quantum}. The Gaussian process is one way to view ANNs and explains why many parameters are not overfitting. ANNs are equivalent to kernels at initialization and throughout the training process. Many parameters in ANNs remain constant, causing them to behave like a kernel \cite{jacot2018neural}. Their values do not depend on the training data, but rather on the architecture of the neural network. 


Machine learning algorithms that use kernels are known as kernel machines. These models operate by mapping data from a low-dimensional space to a high-dimensional one using functions such as Gaussian kernels, which can enhance classification performance. Its inverse process is regularization, which aims to reduce the number of free parameters to prevent overfitting.
Kernel machines are conceptually simpler and more analytically tractable. During training, the evolution of the function represented by an infinite-width neural network mirrors that of a kernel machine. In function space, both models can be visualized as descending a smooth, convex (bowl-shaped) landscape in a high-dimensional space. Due to this structure, it is mathematically straightforward to prove that gradient descent converges to the global minimum. However, the practical relevance of this equivalence remains debated. Real-world neural networks are of finite width, and their parameters can change in complex ways during training. In terms familiar to statistical physicists, this is akin to questioning whether insights from the SK model of spin glasses remain valid in the short-range EA model—i.e., whether the behavior in idealized, solvable models carries over to more realistic, complex systems.

\section{Challenges and perspectives}

ANNs are good at recognizing patterns. However, there is a lack of explainability, bias, hallucination, and transferability in ANNs. More drawbacks that need to be mitigated include reliance on massive datasets, catastrophic forgetting, vulnerability to attacks, high computational costs, and inadequate symbolic reasoning \cite{wang2024hopfield}. In this section, we will elaborate on the challenges primarily relevant to their statistical aspects and provide our perspectives on their solutions.


\subsection{Challenges in ANNs and spin glasses for ANNs}
There are many unanswered questions specifically to ANNs, such as why ANNs work and why backpropagation performs better than other methods in high-dimensional space \cite{belkin2021fit}. We do not know where to start solving them, and if we know, ``a horde of people would do it" \cite{miller2024nobel}. 
To make things worse, ANNs have complicated and diverse structures, which prevent them from constructing a rigorous mathematical foundation. Nonetheless, statistical physics provides essential tools to solve these problems.

Statistical physics is deeply connected to optimization problems and provides approaches to solving them. This connection has been convincingly demonstrated by the fact that simulated annealing provides solutions to combinatorial optimization problems, such as the traveling salesman problem \cite{kirkpatrick1983optimization}. Additionally, the simple “basin-hopping” approach has been applied to atomic and molecular clusters, as well as more complicated hypersurface deformation techniques for crystals and biomolecules \cite{wales1999global}. Breakthroughs in statistical physics are valuable to finding the optimal solutions to ANNs. We anticipate the development of more efficient statistical methods.

Understanding the nature of ground states of short-range spin glasses can provide valuable and indirect insights into the capacity of ANN to memorize patterns.
One crucial direction associated with statistical physics for ANNs is to study the topological structure of non-idealized ANNs and their connection with spin-glass models \cite{mezard2024spin}. We can only connect simplified ANNs with spin glasses, not the more interesting, widely applied ANNs. 
A few challenges exist in understanding spin glasses that have interaction ranges intermediate between those of short-range EA and mean-field SK models, as well as their mapping to general ANNs. Specific examples include (i) finding the global optimal solutions and (ii) whether the ground states in a general spin glass are finite or infinite \cite{newman2022ground}. Numerical and theoretical bottlenecks contribute to the challenges. Numerically, (1) there is limited information on large systems and scaling due to the limited computing resources, and (2) there is a lack of visualization methods to help process high-dimensional data and inspire ideas for analytical solutions. Theoretically, (1) we have no clear picture of the models in the thermodynamic limit, and (2) there is no analytical method to describe short-range spin-glass models. Analytic solutions are limited to the dynamics of models with all-to-all interactions, such as the SK model. It is non-trivial to understand a sparse network where not all neurons or spins interact with each other. However, finding an analytical solution to sparse networks is essential for comprehending general ANNs, just as the SK model is for fully connected ANNs. Recently, Metz proposed a dynamical mean-field method for sparse directed networks and found their exact analytical solutions \cite{metz2025dynamical}.
The general solution is claimed to apply to the study of neural networks and beyond, including ecosystems, epidemic spreading, and synchronization, which represents meaningful progress. However, more efforts are still needed in this direction.







\subsection{Spin-glass physics helps understand ANNs}


Spin-glass physics and ANNs are reciprocal. Important inference problems in machine learning can be formulated as problems in the statistical physics of disordered systems. However, the significant issues we face in analyzing deep networks require the development of a new chapter in spin glass theory \cite{mezard2024spin}. Only a solid theory can transform deep network predictions from best guesses in a black box into interpretable, demonstrable statements whose worst-case behavior can be controlled, although constructing a theory of deep learning is challenging.


We have summarized a few examples where ANNs solve problems in statistical and theoretical physics, such as the phase detection of matter or materials. This is an active research area. Meanwhile, mean-field theory (e.g., cavity method or replica theory) inspires the development of ANN algorithms and enhances our understanding of ANNs, such as the double-descent behavior. Currently, only analytical results are obtained for simple ANNs, and only these simple ANNs are well understood. 
Unlike the previous methods, which are suitable for single-layer or shallow networks, active research is ongoing to develop a statistical mechanical theory for learning in deep architectures, such as the method proposed by Li and Sompolinsky \cite{li2021statistical}.
In the future, it is expected that the properties of more real ANNs can be fully explored.



\begin{figure*} 
\centering
\includegraphics[width=1.0\linewidth]{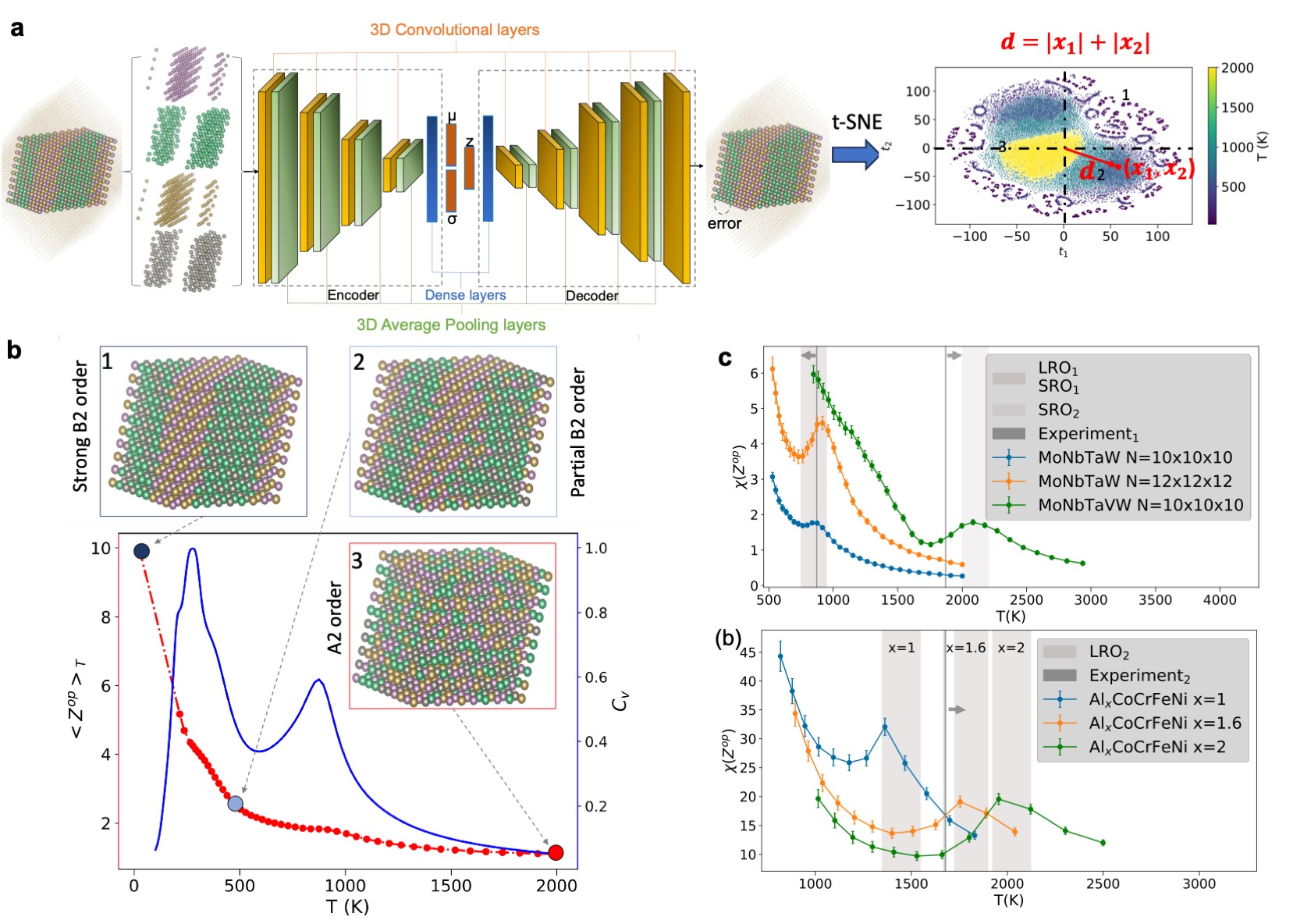}
\caption{Neural network-based order parameter in complex concentrated alloys. {\bf a}, A schematic structure for variational auto-encoder (VAE) was used. The information extracted from the latent space and projected in a two-dimensional (2D) space using t-SNE. The 2D data was then used to construct the order parameter $\langle Z^{op}\rangle_T$ based on Manhattan distance $\sum_i |x_i|$. {\bf b}, The new order parameter describes the different configurations. The first-order derivative of $\langle Z^{op}\rangle_T$ (i.e., $\chi(Z^{op})$) can play a similar role to the specific heat $C_v$, where the peaks represent the two phase transitions. {\bf c}, The temperatures of phase transitions represented by the peak of $\chi(Z^{op})$ are compared with experimental data. This figure is adjusted from Ref. \cite{yin2021neural}.}
\label{fig:order-parameter}
\end{figure*}

\subsection{Do we need new order parameters for ANNs?}

Historically, order parameters play an essential role in describing the thermodynamic behavior of complex systems, as demonstrated by the second-order parameters $\langle q^2\rangle$ proposed by Edwards and Anderson \cite{edwards1975theory}. New order parameters promote the study of glassy systems. Currently, the phase distributions of short-range spin glasses are not fully understood, which may limit our understanding of complex ANNs. We may need new order parameters based on ANNs to describe phase transitions and the phase distribution of spin glasses.  Developing a rigorous mathematical description of metastable states and solutions appears to be necessary. In a previous study, we adopted a similar idea and proposed a neural-network-based order parameter to study the complex concentrated alloys, where multiple principal elements co-exist that introduce maximal disorder (Figure \ref{fig:order-parameter}) \cite{yin2021neural}. We successfully utilized the new order parameter to differentiate between different phases. We found that the predicted phase transition temperatures are consistent with experimental results.
Similarly, due to the complex feature of ANN phase transitions, we may need one ANN to help us understand the phases of another ANN. ANNs can capture size-independent patterns that pave the way to understand the ground states of spin glasses at the thermodynamic limit. Currently, only spin-glass results of relatively small sizes are available, limited by computing resources (speed of supercomputers). However, we still need to overcome difficulties, including feature extraction and feature synthesis used in ANNs.


\subsection{Quantum computing for spin glasses and ANNs}

Due to the rugged energy landscapes of spin glasses, even today's most powerful supercomputers cannot model the large size of this complex system, leaving the behavior of these systems largely unexplored. Quantum computers can mitigate this issue with their exponential scaling. 
The breakthroughs in both hardware and algorithms in quantum computing provide new opportunities.
A team from Google reported an error-corrected chip, Willow, featuring 105 qubits based on superconducting qubits, in December 2024 \cite{Acharya2025}. The performance of the Willow chip is challenged by China's Zuchongzhi 3.0 processor \cite{gao2025establishing}, a superconducting quantum computer prototype featuring 105 qubits, which was reported in March 2025.
The hope of quantum computing based on Majorana particles, a type of quasi-particle in topological insulators, rose in February 2025. Microsoft published its intermediate, controversial result on its quantum computer, Majorana 1 \cite{quantum2025interferometric}.
Quantum computers can be used to efficiently explore the phase spaces of spin-glass-like systems. Simulating Ising spin glasses on a quantum computer provides new opportunities \cite{Lidar1997}. For example, the Ising-type Sachdev-Ye-Kitaev (SYK) model that describes the dynamics of wormholes between black holes for fermions can be solved numerically by quantum computing \cite{gao2021traversable,jafferis2022traversable}. Quantum computing can also be applied in quantum machine learning, such as in the development of quantum versions of ANNs or quantum neural networks \cite{boreiri2025topologically}. This will become more promising with the emergence of new quantum algorithms. Recently, King and colleagues used quantum annealing processors to simulate quantum dynamics in programmable spin glasses \cite{King2025-D-Wave}, one application of quantum computers of practical interest. Given the topological relationship between spin glasses and ANNs, this will provide valuable insights into understanding ANNs.

Additionally, it is anticipated that ANNs can leverage the benefits of quantum computing in terms of efficiency and accuracy. For example, the quantum version of the Hopfield network was developed by replacing classical Hebbian learning with a quantum Hebbian learning rule \cite{rebentrost2018quantum}. The researchers demonstrated the ability to store exponentially large networks with a polynomial number of qubits by encoding them as amplitudes of quantum states. Their quantum algorithm achieves a quantum computational complexity that is logarithmic in the dimensionality of the data.
The advantages of quantum ANNs, as well as other quantum algorithms, are questionable and unknown in near-term quantum computers. To address this critical question, Abbas {\it et al.} proposed the so-called effective dimension to measure the power and trainability of quantum ANNs \cite{abbas2021power}. Assisted by the measure, they showed a quantum advantage of a class of quantum ANNs.
Quantum ANNs are currently under active investigation and are still in their early stages of development. 

\subsection{More opportunities}

One exciting future direction of ANN is neural computing, also known as synaptic computation \cite{abbott2004synaptic,wang2018fully,weilenmann2024single,zhou2024field}. The storage and computing of the current ANNs are separate. Differently, biological neurons store data and perform computations through the same neural network, making it highly efficient. Since associative ANNs have demonstrated their capability for memory storage and have been successfully applied in computation, it is natural to take the next step and design memristive neural networks. Memristors can store and compute information simultaneously. A memristor is a combination of a memory unit and a resistor, which can mimic biological synaptic functions in ANNs.
The resistance of a memristor changes based on the history of applied current or voltage. A prominent advantage of memristors is that they are non-volatile and can retain memory in the absence of power. For example, Wang {\it et al.} developed fully memristive neural networks for pattern classification with unsupervised learning \cite{wang2018fully}. 
In a recent study, Weilenmann and colleagues explored energy-efficient neural networks using a single neuromorphic memristor that mimics multiple synaptic mechanisms \cite{weilenmann2024single}. More details can be found in a recent review of the memristive Hopfield neural networks applied to chaotic systems \cite{lin2023review}.
Memristors are artificial analogs of biological neurons that harness computing power by mimicking their structure and function. Recently, another exciting direction that deserves more attention is the direct participation of neurons {\it in vitro} in computational processes. These studies integrate adaptive {\it in vitro} neurons and {\it in silico} high-density multi-electrode arrays into digital systems to perform computations \cite{kagan2022vitro}. Nonetheless, these research directions are still in their infancy and offer a lot of opportunities to explore the potential of neural computing. Since they are not the focus of this article, we will not discuss them in depth.


\section{Conclusions}
We have reviewed the applications and history of artificial neural networks, as well as their connections with biology and statistical physics, particularly in the context of spin glasses. We showed the deep connections of these multidisciplinary directions. For example, we demonstrate how the replica theory is applied to understand the behavior of artificial neural networks. We also discussed the challenges and possible solutions. One of the significant problems is understanding and also accelerating the training of large artificial neural networks, which can benefit from the integration with quantum computing and neural computing. Quantum computing can provide exponential acceleration, while neural computing can mitigate the storage limit, reducing the latency in data transfer. Theoretically, the complex behavior, reliability, and stability of ANNs require more research efforts, which lag behind their applications.
Arguably, the most challenging problems involve understanding biological and artificial neurons. The former involves the formation of consciousness, while the latter involves the realization of artificial intelligence. We concluded that statistical physics bridges the gap between these key multidisciplinary problems and will provide valuable methods to find their answers.

\noindent








\bibliographystyle{plain}
\bibliography{scibib}



\end{document}